\documentclass{article}

\usepackage{arxiv}

\usepackage[utf8]{inputenc} 
\usepackage[T1]{fontenc}    
\usepackage{hyperref}       
\usepackage{url}            
\usepackage{booktabs}       
\usepackage{amsfonts}       
\usepackage{nicefrac}       
\usepackage{microtype}      
\usepackage{lipsum}
\usepackage{graphicx,verbatim,float,subfigure,graphicx,color,siunitx}
\usepackage{amsmath}
\graphicspath{ {./images/} }

\usepackage{amssymb}
\usepackage{svg}
\usepackage{multirow}  
\usepackage[export]{adjustbox}
\usepackage{soul}
\usepackage{textcomp}
\usepackage{multirow}
\usepackage{enumitem}
\usepackage[sort&compress,numbers]{natbib}

\title{Moisture-Driven Degradation Mechanisms in the Viscoelastic Properties of TPU-Based Syntactic Foams}

\author{
 S. P Subramaniyan and  P. Prabhakar$^*$ \\
  Dept. of Mechanical Engineering \\
  University of Wisconsin-Madison \\
  Madison, WI 53706  \vspace{0.05in} \\
  \texttt{$^*$pavana.prabhakar@wisc.edu}
}

\begin{document}
\maketitle
 
\newcommand{\SPS}[1]{\textcolor{red}{\bf{Sabari: #1}}}

\begin{abstract}
Syntactic foams have found widespread usage in various applications including, marine, aerospace, automotive, pipe insulation, electrical cable sheathing, and shoe insoles.  However, syntactic foams are often exposed to moisture when used in these applications that  potentially alter their viscoelastic properties, which influences their long-term durability. Despite their significance, previous research has mainly focused on experimental studies concerning mechanical property changes resulting from filler loading and different matrix materials, overlooking the fundamental mechanisms resulting from moisture exposure. The current paper aims to bridge this gap in knowledge by elucidating the impact of long-term moisture exposure on TPU and TPU-based syntactic foam through multi-scale materials characterization approaches. Here, we choose a flexible syntactic foam manufactured using thermoplastic polyurethane elastomer (TPU) reinforced with glass microballoons (GMB) through selective laser sintering. Specifically, the research investigates the influence of moisture exposure time and the volume fraction of GMB on chemical and microphase morphological changes, along with their associated mechanisms. The study further examines how these microphase morphological changes manifest in viscoelastic properties.
\end{abstract}



\section{Introduction}\label{intro}

Thermoplastic polyurethane elastomers (TPUs) are block copolymers with alternating hard and soft segments. The soft segments are typically derived from polyester or polyether (polyol), while the hard segments are produced by combining a chain extender with an isocyanate. The mechanical properties of TPUs, such as modulus, strength, hardness, damping, and tribological performance, can be customized by varying the formulation and constitution, as well as by incorporating additives.
TPUs have found widespread use in numerous applications, including sealings, hoses, shoe soles, cable sheaths, films, foams, and automotive interiors. They exhibit the processing abilities of plastics and high elasticity of rubber, making them a versatile and sought-after material in the industry. Furthermore, TPUs are characterized by low-temperature flexibility, excellent abrasion resistance, good processing characteristics, and biocompatibility, which make them ideal for use in diverse fields, such as transportation, construction, and biomedical materials.\cite{Aurilia2011,Bruckmoser2014,Deng1994,Osswald2012,Xu2021}.

Extensive research has been performed on the influence of hard-to-soft segment ratio, hard segment type, chain extender type, and segment length on the Hydrogen bonding, crystallization behavior\cite{Xiu1992}, mechanical performance\cite{Xiang2017, Walo2014,Petcharoen2013} of TPU. Despite the vast amount of research conducted on thermoplastic polyurethane elastomer, there remains a significant gap in understanding the long-term effects of moisture-induced microphase morphology changes facilitated by hydrolysis, particularly with regard to their impact on viscoelastic properties. Our current paper tries to bridge the gap mentioned above by correlating the microphase morphology and its properties.

Yang et al.\cite{Yang2006} delved into the impact of moisture on the thermomechanical properties of an ether-based polyurethane shape memory polymer. They indicated that the glass transition temperature of the polymer dropped considerably after water immersion, resulting in weakened hydrogen bonding between the N-H and C=O groups. The absorbed water was divided into two categories: free and bound water, with bound water having a significant impact on the uniaxial tensile behavior. The researchers further examined the recovery stress and recovery strain brought about by water and compared them with those induced thermally. They discovered that by heating the polymer to 180°C, all absorbed water could be removed, and the interactions between the polymer chains and water were eliminated. 

Bourbakri et al.\cite{Boubakri2009} investigated the influence of a hygrothermal environment on the mechanical and thermal properties of thermoplastic polyurethane elastomer (TPU). They determined the diffusion coefficient using Fick's diffusion model and showed the plasticization mechanism using differential scanning calorimetry (DSC). They concluded that polymer degradation was irreversible, and some of the reversible physical properties after moisture aging were also accessed. They found that the wear resistance decreased after moisture exposure. Subsequent work by Bourbakri et al.\cite{Boubakri2010} investigated the effects of aging factors on the mechanical properties of TPU material. They concluded that the diffusion of water molecules largely depends on aging temperature and affects mechanical response, with deterioration in properties strongly influenced by immersion time and temperature. The storage modulus decreased with aging temperature, and degradation was manifested by small voids and deterioration under severe conditions.

The influence of thermal treatment on the permeation properties of a polyester-based thermoplastic polyurethane (TPU) was investigated by Puentes-Parodi et al.\cite{Puentes-Parodi2019}. Annealing the polymer at specific conditions (100 \textdegree C/20 h) increased diffusion and permeation rates. The higher diffusion rates of gases and water molecules after annealing are likely attributed to a relaxation process that redistributes crystalline and amorphous domains, increasing the free volume within the macromolecular architecture of the TPU. Indirect characterization methods such as mechanical, DSC, FTIR, water immersion, and permeation tests were used to draw these conclusions.
Bardin et al.\cite{Bardin2020} developed a hydrolytic kinetic model to predict chain scission in both stabilized and unstabilized thermoplastic polyurethane elastomers. They also investigated the structure-failure property relationship between molar mass and elongation at break.
In their study, Xu et al.\cite{Xu2021} investigated into the impact of water on the structure and mechanical properties of thermoplastic polyurethane (TPU) blends that had varying ratios of high and low hardness TPU compositions. They found that the wet TPU samples underwent softening and hardening due to changes in the structure at the interface between the hard and soft domains in TPU. This was a combined effect that included the plasticization of free water, breakage and/or weakening of the original hydrogen bond by water, and the formation of new hydrogen bonds (elastic active cross-linking points) by free carbonyl groups in TPU and water. Mishra et al.\cite{mishra2015long} investigated the hydrolysis stability of several commercial-grade TPUs exposed to phosphate buffer solution for a prolonged period. They found a substantial reduction in molecular weight and ultimate tensile strength of the material. They also found the rate of reduction in the molecular weight to the temperature follows second-order kinetics. Choi et al.\cite{choi2023degradation} investigated the degradation kinetics and predicted the lifetime of TPU encapsulants exposed to seawater. They considered a 50\% reduction of initial tensile strength as a failure criterion. They estimated that the TPU encapsulants must be replaced within at least 27.31 years to prevent material system failure.


Syntactic foams are a type of closed-cell foam that consists of hollow microspheres embedded in a matrix material. The microspheres are typically made of glass, cenosphere (fly ash waste material), or metal, and the matrix materials are made of polymers, ceramics, or metal. Polymer syntactic foams are often used in marine, automotive, aerospace, and electrical applications due to their high strength, low density, high buoyancy, low thermal expansion, vibrational damping, and acoustic, thermal, and electrical insulation. These applications include energy-absorbing cores in sandwich composites, radome material, acoustic insulation for underwater sonar devices, thermal insulation, and electrical insulation cable insulation. The properties of syntactic foams can be tailored by changing the matrix type, hollow microspheres volume fraction, and hollow microspheres wall thickness.\cite{Gupta2014,Afolabi2020,Shahapurkar2018}. Polymers used in syntactic foam can be thermoset (epoxy, phenolic, cyanate ester, polyurethane), thermoplastic (polyethylene, nylon, polystyrene), or elastomers. This work uses Thermoplastic polyurethane elastomer as the matrix material with soda-lime borosilicate-based glass as microballoon reinforcements.

Researchers have also developed elastomeric and rubber-based syntactic foams, which find application in various sectors such as shoe soles, pneumatic tires, wires, and cable compounds. Thermoplastic polyolefin elastomers-based syntactic foam has also been used for buoyancy and coaxial cable insulation\cite{amos2015hollow}.
Early studies have shown that incorporating glass microbubbles (GMBs) into rigid polyurethane foam using commercial molding processes significantly improves the compressive strength, flexural strength, moduli, and impact resistance of the materials without increasing their density\cite{Barber1977}. In particular, incorporating 10\% GMB by weight resulted in a 70\% improvement in the compressive modulus\cite{Hagarman1985}. Polyurethane elastomers (TPUs) for underwater applications were developed by Hyungu Im et al\cite{Im2011}. The TPUs were prepared using poly(tetramethylene glycol) (PTMG) and methyl diphenyl diisocyanate (MDI). Hollow glass microspheres (HGMs) were grafted onto the TPU matrix to improve the interfacial adhesion between the TPU and the HGMs. The tensile strength of the composite increased with increasing HGM content, while the swelling ratio and density of the composite decreased with increasing HGM content. The TPU-HGM composites also exhibited enhanced mechanical strength as compared to TPU after being immersed in seawater and paraffin oil. The study by Tewani et al.\cite{Tewani2022} investigated the effect of print parameters and GMB particle size on the mechanical characteristics of selective laser-sintered TPU-based syntactic foam. The outcomes indicated that the particle size of GMBs had a significant impact on the microscale architecture of the syntactic foam. Specifically, the smaller GMBs were found to be entirely embedded within the segregated matrix of TPU, whereas the larger GMBs were lodged between the pores of the segregated matrix. This particular distribution pattern of GMBs in the TPU-based syntactic foam resulted in distinctive mechanical behavior, making it possible to tailor the material at microscale and macroscale levels.

In the current study, Thermoplastic Polyurethane Elastomer (TPU) reinforced with Glass MicroBalloons (GMB) was manufactured using Selective Laser Sintering. 
Further, the current work aimed to investigate the effect of moisture, GMB volume fractions, and temperature on the material's viscoelastic properties. The chemical changes resulting from moisture aging were also studied and correlated with the observed material properties.
\section{Motivation}
The core objective of this work is to elucidate the influence of internal architecture and moisture-induced degradation in the functional properties of additively manufactured TPU-based syntactic foams.

The fundamental research questions this work addresses are as follows:
\begin{enumerate}
    \item  How does long-term moisture aging alter the microphase morphology and chemical structure of TPU and TPU-based syntactic foams? How do these changes manifest in terms of viscoelastic properties?
    \item How do GMBs alter the moisture-induced degradation kinetics of TPU and TPU-based syntactic foams?
\end{enumerate}

\begin{figure}[h!]
 \centering
  \includegraphics[width=14cm]{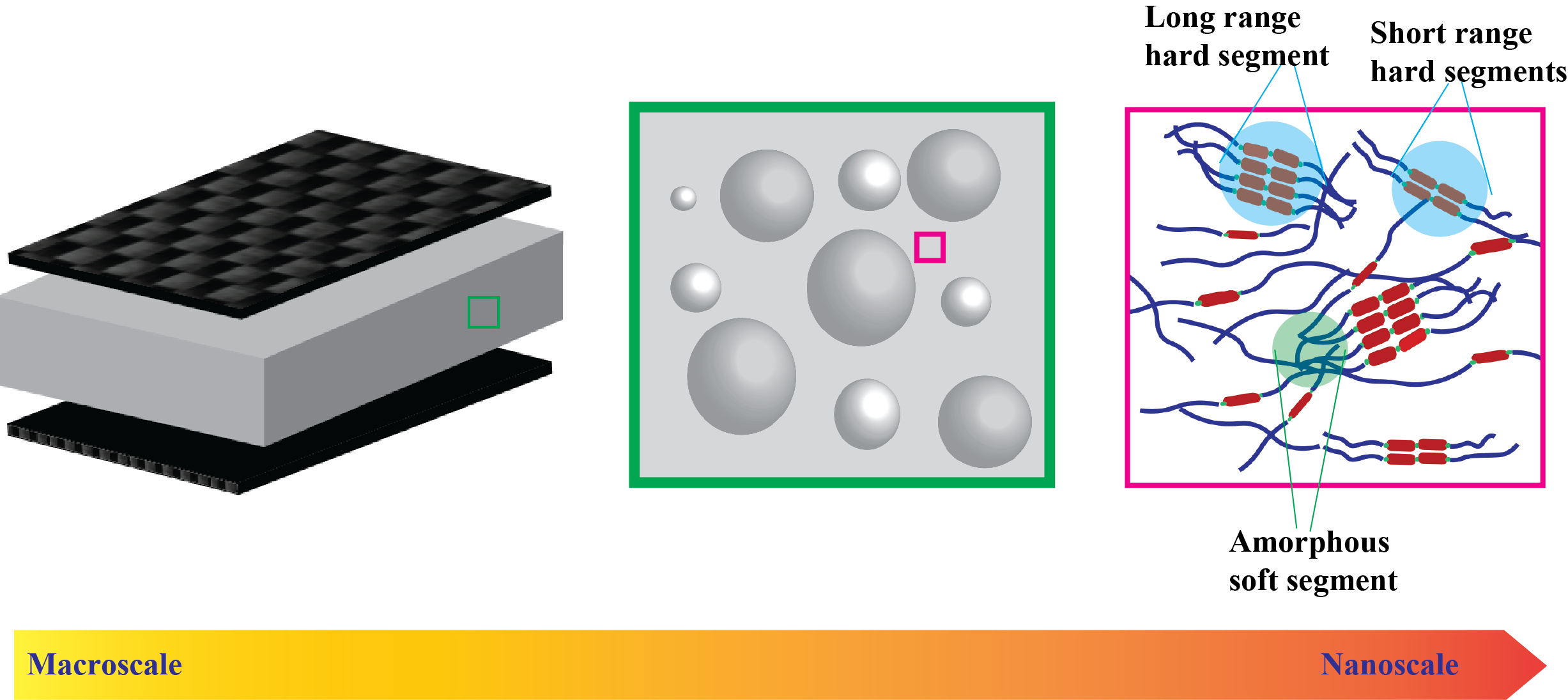}
  \caption{Illustration of hollow glass microballoon (GMB) reinforced Thermoplastic Polyurethane (TPU) elastomer at different length scale}\label{fig:material}
\end{figure}

\section{Methodology}\label{method}

\section{Materials}

The multiscale material hierarchy of TPU-based syntactic foam is depicted in Fig~\ref{fig:material}. At the macroscale level, the TPU-based syntactic foam is a lightweight foam structure that can be designed into various architectures using additive manufacturing techniques to control the macroscale property\cite{Tewani2022}. At the microscale level, glass micro balloons are reinforced into a TPU matrix, where the volume fraction can be adapted to achieve the desired density and property of interest. TPU is a block copolymer of hard and soft segments at the nanoscale level. The hard segments are obtained by reacting diisocyanates with a diol or diamine chain extender, while the soft segments are made from amorphous polyester or polyether. The thermodynamic incompatibility between the soft and hard segments might cause macrophase separation, commonly found in polymer blends, and can affect the physical properties. However, in TPU, covalent bonds between hard and soft segments prohibit the formation of macrophase separation, thereby creating different microphase-separated morphology. Although these structures are observed at the nanoscale, we will adhere to the older terminology of microphase morphology. The modifications in the microphase morphology play a crucial role in determining the viscoelastic properties \cite{yilgor2015critical,cheng2022review}.

\subsection{Manufacturing and conditioning}
This work considers syntactic foams composed of Thermoplastic polyurethane elastomer(FLEXA Grey - Sinterit) matrix and hollow micro-glass balloon(K20 - 3M) fillers additively manufactured using a selective laser sintering process. The composite powder preparation involves mixing TPU with GMB at different volume fractions (20\% and 40\%) using an automated mixer at 30 V for 5 minutes, followed by 70 V for 3 minutes to achieve dispersion of GMB in TPU. Mixed powder is fed into SLS equipment (Lisa-pro). The sintering process is carried out with an input laser power ratio of 1.5 and layer height of 0.075 mm. All the input parameters are chosen based on a parametric study performed by Tewani et al.\cite{Tewani2022}.

Pristine TPU and TPU with 20\% and 40\% GMBs are manufactured and immersed in de-ionized water at 23 \textdegree C. Then, the samples are removed and desorbed at 50 \textdegree C in an oven for 24 hours to remove the free water in the samples. The primary focus of this work is to understand the underlying mechanism of bound water's influence on chemical, thermal, and mechanical properties.

\subsection{Chemical characterization}
Fourier Transform Infrared spectroscopy (FTIR) is performed at room temperature(~23\textdegree C) to analyze the chemical changes associated with the addition GMBs and moisture aging in TPU-based syntactic foam. Spectrometer FTIR Nicolet is50R with Attenuated Total Reflectance (ATR) is used to measure the absorbance spectrum from 4000 $cm^{-1}$ to 600 $cm^{-1}$ wavenumber at a rate of 4 $cm^{-1}$ for each scan and averaged using 64 scans.

\subsection{Thermal stability characterization}
Differential Scanning calorimetry (DSC), using TA equipment QA 200,  is carried out to comprehend the effects of moisture and GMB volume fraction on the thermal stability of TPU-based syntactic foam. We employ a sample weight of 4–8 mg in a controlled inert atmosphere to ignore the complexity of oxygen-induced breakdown. By heating from -50\textdegree C to 225 \textdegree C and cooling back down to -50 \textdegree C, we undertake dynamic analysis at a constant temperature ramp rate of 10 \textdegree C/min. Calculating melting enthalpy ($\Delta H_{T}$) involves using the second heating curve.

\subsection{Viscoelastic characterization}
Dynamic Mechanical Analysis is performed on TPU-based syntactic foam using RSA -III ( TA Instruments). The influence of GMB volume fraction, temperature, and moisture on the viscoelastic properties of syntactic foam is investigated. A tensile test is opted to determine the viscoelastic properties since it provides insight into adhesion properties between TPU and untreated GMB and other viscoelastic properties. A 30 x 10 x 1 $mm^3$ sample geometry is opted and heated from -40\textdegree C to 120\textdegree C at a strain of 0.05\%  and frequency of 1 Hz. A soak time of 1 minute is chosen at each temperature step to reach the thermal equilibrium in the sample.


\section{Results}\label{res}

This study investigated the influence of moisture on TPU (thermoplastic polyurethane) and TPU-based syntactic foam.  In Section~\ref{chemical_change}, the chemical characterization of unaged and aged samples is performed using FTIR. This analysis determines the degree of phase separation and weight fractions of hard and soft domains and the presence of mixed phases by assessing the intensity of carbonyl group bondings. Section~\ref{DSC-thermal_characterization} uses DSC to characterize the material's thermal behavior. The study involves deconvoluting the endothermic peaks and evaluating melting enthalpies to determine different microphase morphology changes. Section~\ref{viscoelastic_characterization} discusses the viscoelastic properties of unaged and degraded samples, including determining the degree of adhesion deterioration in the TPU-GMB interface.

\subsection{Influence of Moisture and GMB Volume Fraction on Chemical Properties of Syntactic Foam}\label{chemical_change}

\begin{figure}[h!]
\centering
\subfigure[]{
\includegraphics[width=7.5cm]{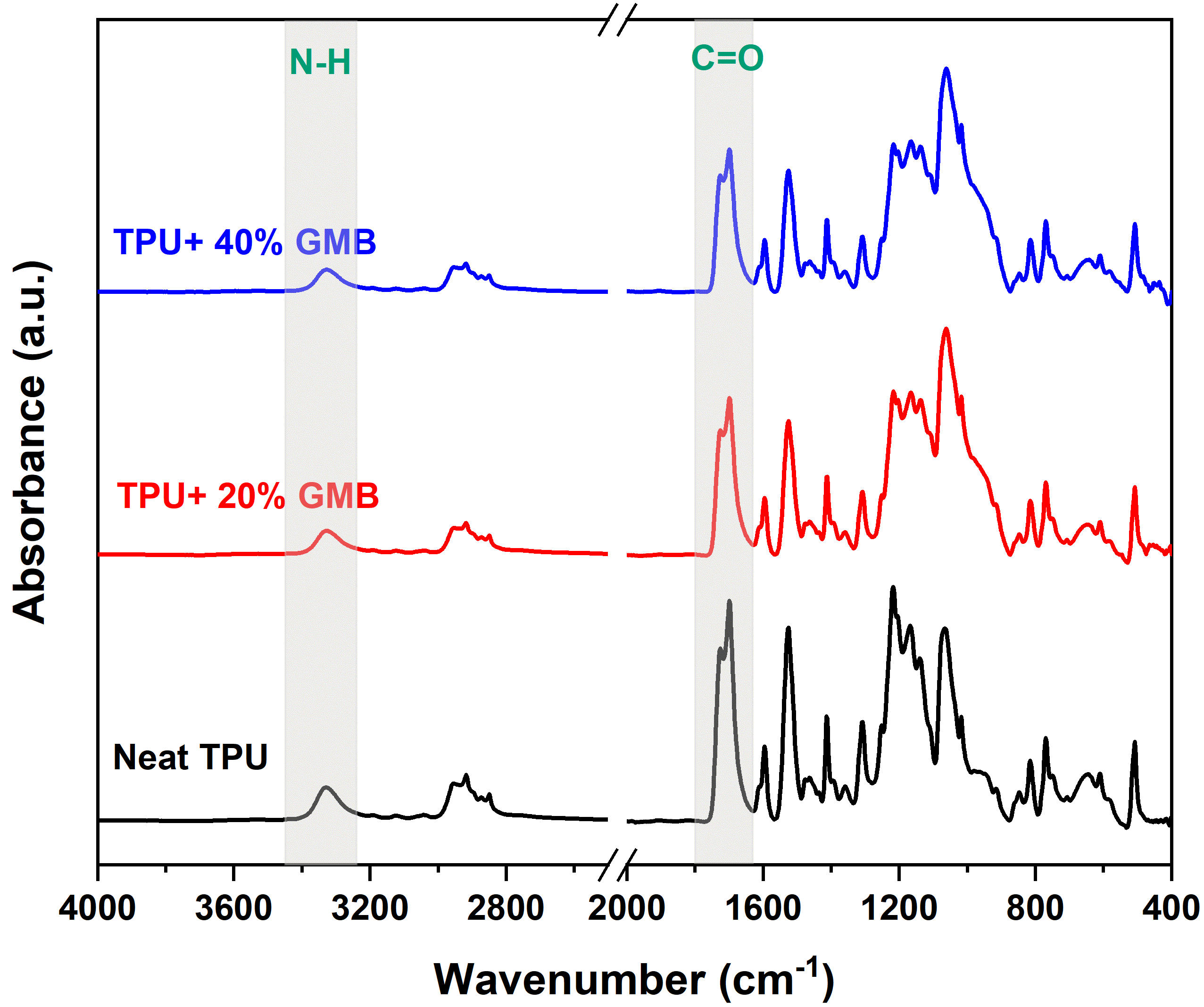}\label{FTIR_spectra}
}
\hspace{0.1in}
\centering
\subfigure[]{
\includegraphics[width=7cm]{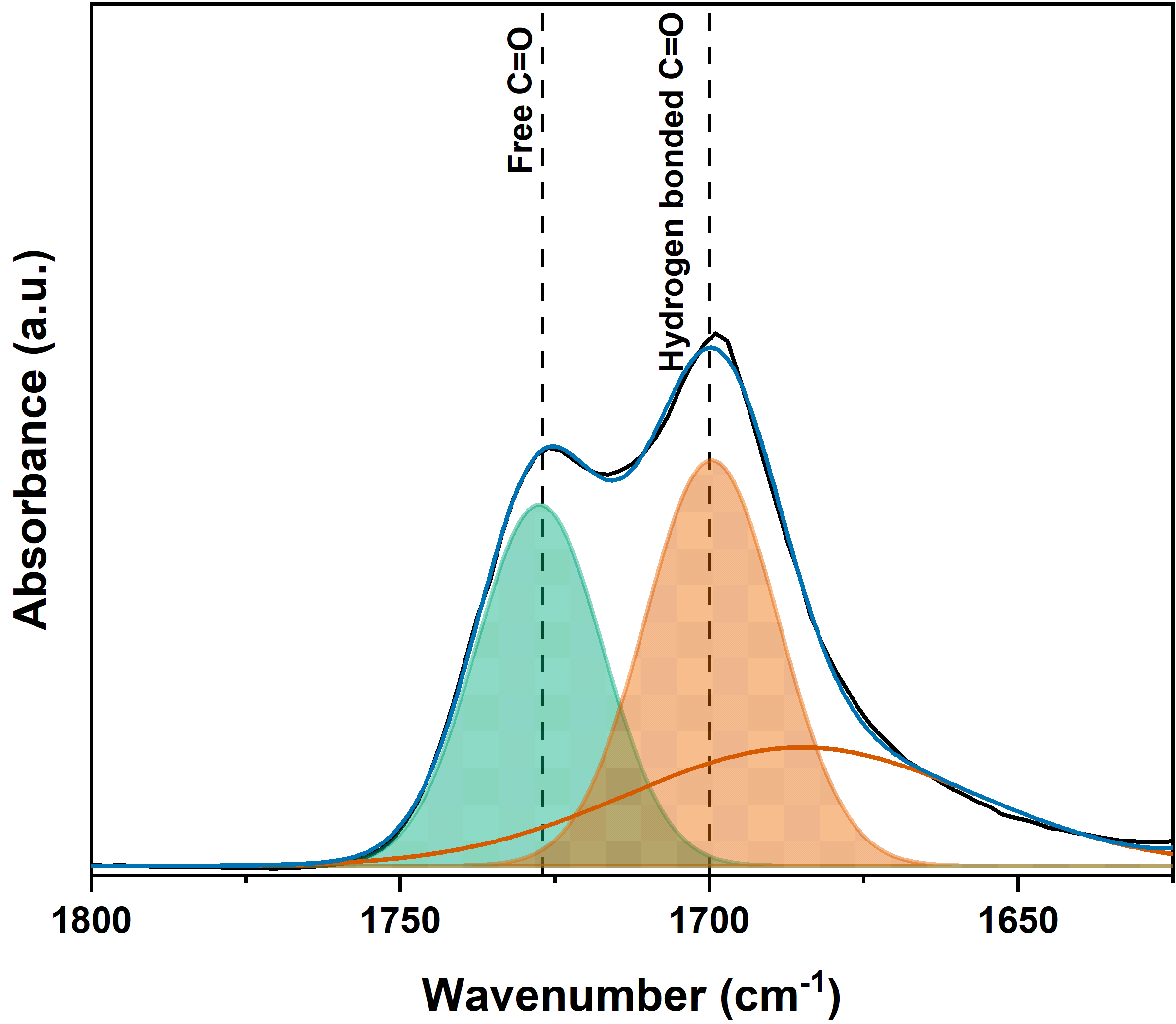} \label{Deg_of_phase_separation}
}
\caption{(a) FTIR spectra of TPU and TPU-based syntactic foam and (b) Schematic illustration of Gaussian peak deconvolution of hydrogen-bonded carbonyl group to non-hydrogen bonded carbonyl group}
\end{figure}

FTIR data obtained for pristine TPU and TPU-based foams is analyzed, followed by establishing the influence of moisture exposure on them. From Fig~\ref{FTIR_spectra}, the spectral peak found at $\sim$3327 $cm^{-1}$ is attributed to the stretching vibration of the N-H group in polyurethane. There is no visible peak around $\sim$3495 $cm^{-1}$ associated with free N-H, which indicates that all the bonds are completely consumed during the processing. The peaks found between 3000 $cm^{-1}$ to 2800 $cm^{-1}$ are attributed to the stretching of $CH_2$ groups, and the peaks located at $\sim$1725 $cm^{-1}$ and $\sim$1699 $cm^{-1}$ are attributed to the stretching vibration of free carbonyl and hydrogen bonded carbonyl groups, respectively \cite{Bruckmoser2014}. The intensity of all the peaks between 1000 $cm^{-1}$ to 1200 $cm^{-1}$ is increased for the sample with 20\% and 40\% GMB volume fraction primarily due to the stretching of the Si-O-Si bond of GMB around 1100 $cm^{-1}$\cite{Im2011}. The peak observed at $\sim$1166  $cm^{-1}$ signifies the stretching vibration of the ester group, which is found within the soft segment.

\begin{figure}[h!]
 \centering
  \includegraphics[width=12cm]{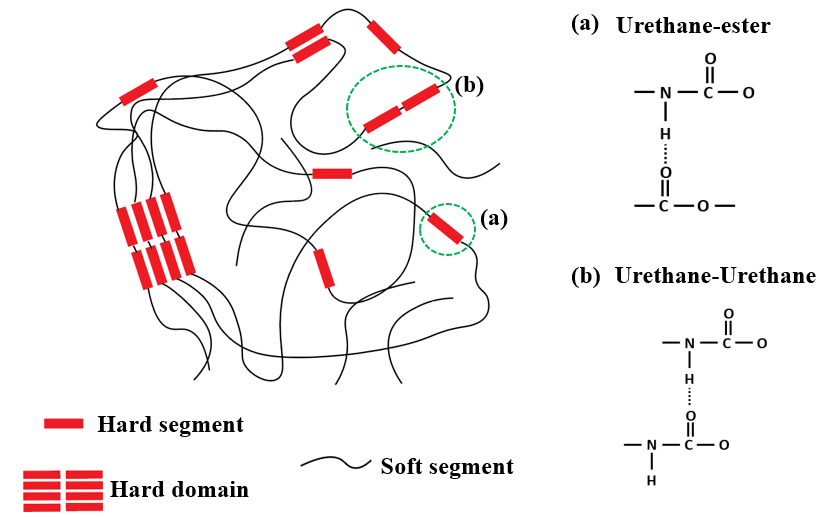}
  \caption{Schematic representation of  hydrogen bonding in thermoplastic polyurethane elastomer}\label{Bond_type}
\end{figure}

Fig~\ref{Bond_type} depicts the hydrogen bonding interactions established within the thermoplastic polyurethane elastomer containing a polyester soft segment; the N-H groups present in the urethane hard segment act as proton donors, while the carbonyl group present in urethane hard segment and polyester soft segment acts as a proton acceptor. Thus, the inter-urethane bond has a strong, cohesive force due to the attraction of highly polar hard segments, forming a hard segment-rich hard domain. On the other hand, the N-H and polyester bonding indicates the cohesion of the hard segment with the soft segment\cite{Xiu1992}. Degree of Phase Separation (DPS), which is the fraction of the hydrogen-bonded carbonyl group, is determined by the ratio of the Area of absorbance peak at 1699 $cm^{-1}$ to the Total Area of absorbance peak at 1699 and 1725 $cm^{-1}$ shown in Fig~\ref{Deg_of_phase_separation}. The Degree of Phase Mixing (DPM) is determined from DPS as 1-DPS. 

\begin{equation}\label{DPS}
    DPS\hspace{1 mm}  = \frac{A_{C=O,bonded}}{(A_{C=O,bonded}+A_{C=O,free})}
\end{equation}

\begin{equation}\label{DPM}
    DPM = 1 - DPS
\end{equation}

\begin{figure}[h!]
 \centering
  \includegraphics[width=7cm]{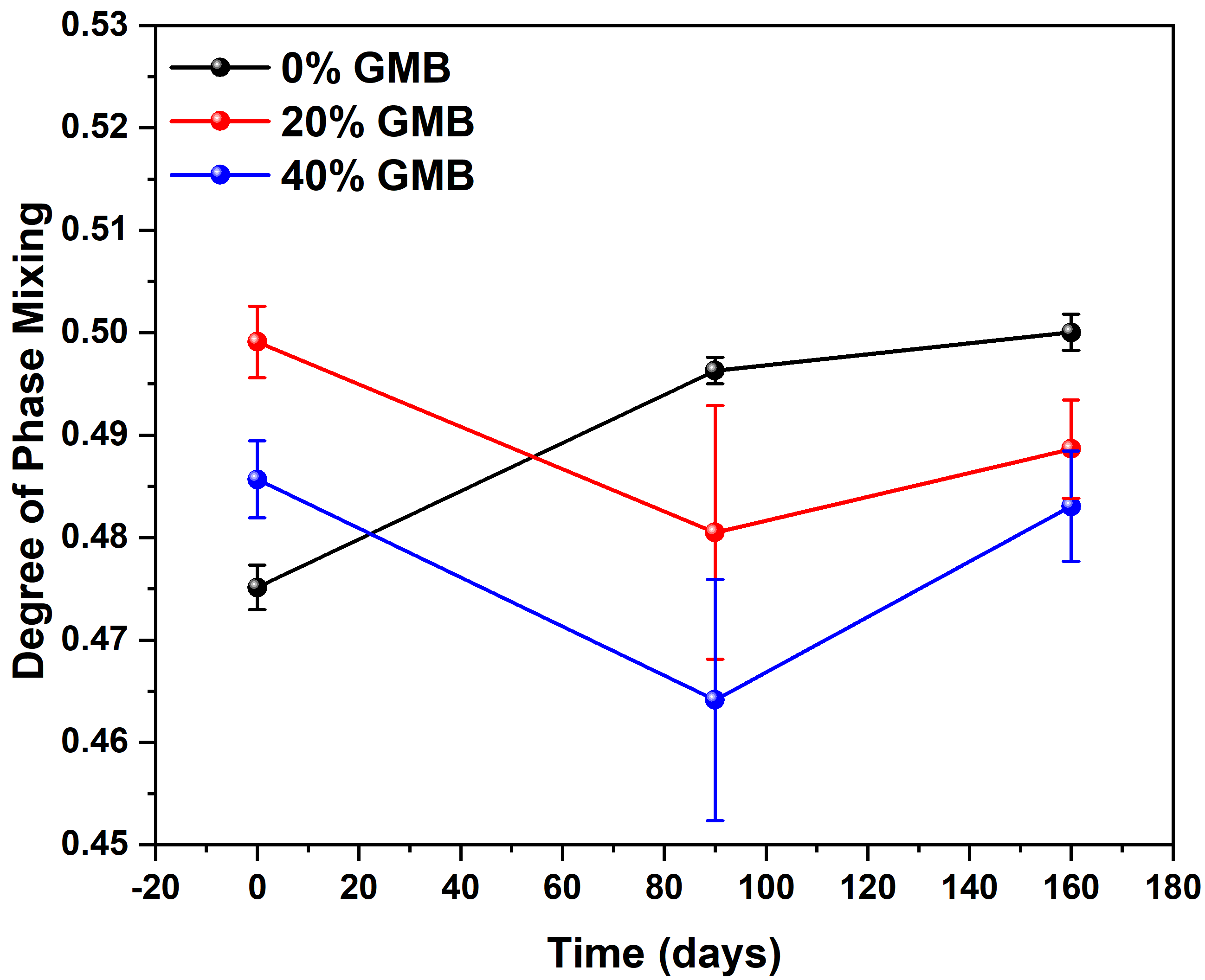}\label{Deg_Phase_mix}
  \caption{Degree of phase mixing as a function of moisture exposure time and GMB volume fraction}
\end{figure}

Upon analyzing the unaged samples, it was observed that TPU with 20\% GMB exhibited a higher degree of phase mixing, followed by TPU with 40\% GMB and neat TPU. However, after 90 days of moisture exposure, neat TPU displayed an increased degree of phase mixing, while TPU with 20\% GMB and TPU with 40\% GMB exhibited a decrease. Furthermore, after 160 days of moisture exposure, the degree of phase mixing in TPU with GMBs experienced a significant increase, whereas it only slightly increased in neat TPU. One primary reason for this observation is that after 90 days of moisture exposure, water molecules absorbed by the samples may form bridges between NH---C=O, weakening the original hydrogen bonds. Additionally, there is a possibility of water molecule bridges forming between two C=O groups. The interaction between water molecules and the polymer diminishes upon desorption, and the original hydrogen bonding is restored\cite{Yang2006}. However, after 160 days, ester hydrolysis becomes dominant, reducing the length of the soft segments through chain scission. As a result, the phase mixing for TPU with syntactic foam increases, indicating a decrease in phase separation. This finding is supported by the work of Jouibari et al.\cite{sahebi2019novel}, who also observed an increase in the degree of phase separation with longer soft segments. It should be noted that although FTIR results demonstrate different types of chemical bonding between hard-hard and hard-soft segments, it may be challenging to conclude the formation of the hard domain using DPS and DPM techniques. This is because the polyester carbonyl forms hydrogen bonding with hard segments, which could lead to misinterpretation of the data\cite{valim2022unraveling}.

\subsection{Influence of moisture and GMB volume fraction on the thermal stability of syntactic foam}\label{DSC-thermal_characterization}

\begin{figure}[h!]
\centering
\subfigure[]{
\includegraphics[width=6.8cm]{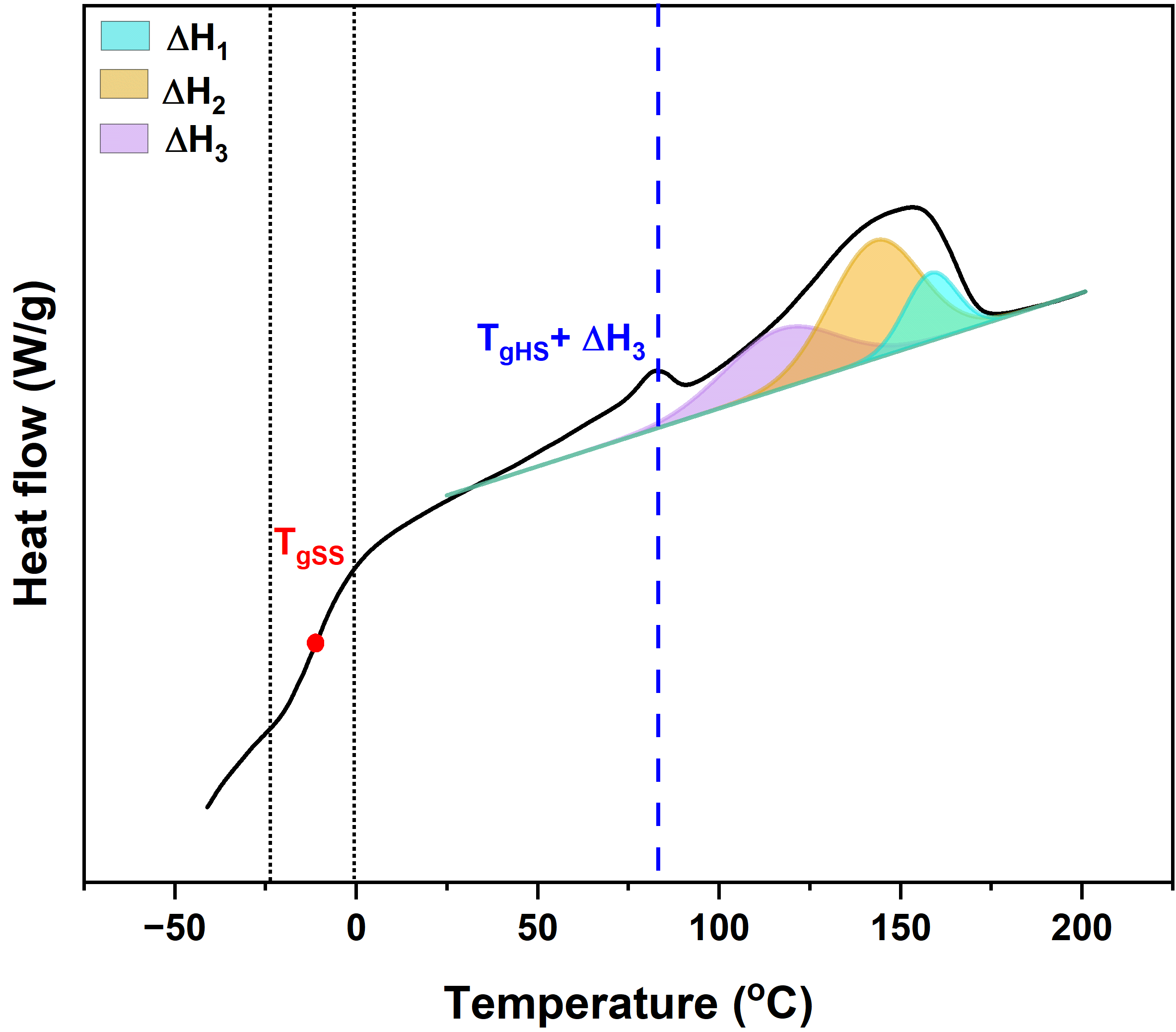}\label{DSC_peaks}
}
\hspace{0.2in}
\centering
\subfigure[]{
\includegraphics[width=7.25cm]{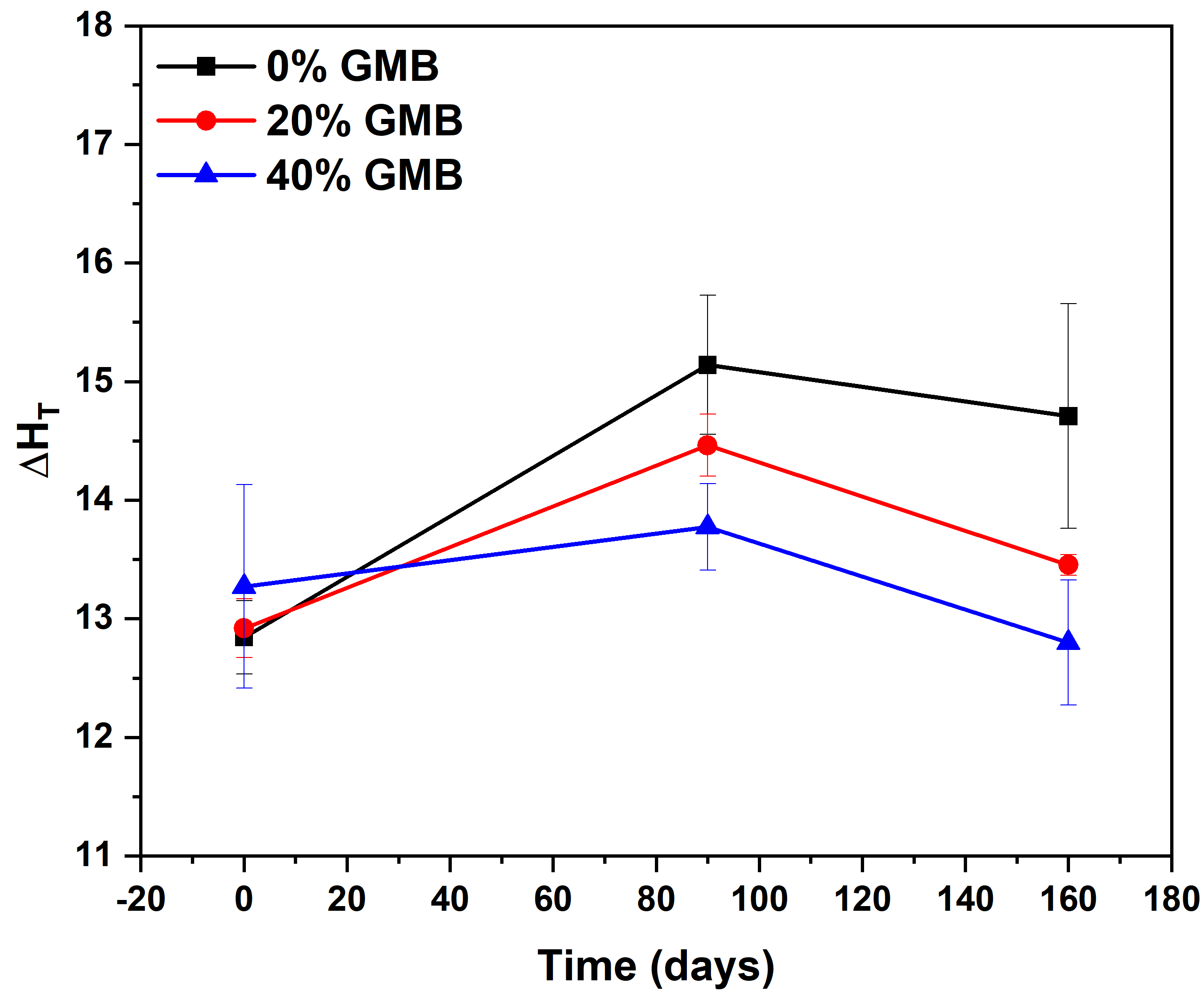}\label{enthalpy}
}
\caption{(a)Endothermic heating cycle of TPU with hidden peaks that define different hard segment morphologies (b) Total melting enthalpy of for different volume fraction of GMB as a function of moisture exposure time.}
\end{figure}

Moisture-induced microphase morphology changes in TPU and TPU-based syntactic foam can be effectively captured using differential scanning calorimetry (DSC). The second heating curve of the samples is typically used for analysis, as the first heating curve often reflects the thermal history of the material. The endothermic cycle, which consists of a weak transition in the sub-zero degree temperature range (as shown in Fig~\ref{DSC_peaks}), is attributed to the glass transition temperature of the soft segment($T_gSS$). This study quantifies the glass transition temperature using dynamic mechanical analysis (DMA), which will be discussed later. Fig~\ref{DSC_peaks} illustrates that upon heating, the hydrogen bonds between hard and soft segments, and between hard segments, break and dissociate. This is followed by melting hard segments involving different transitions and multiple endothermic peaks. While heating the sample, a weak transition accompanied by a small endothermic peak appears at around 80 \textdegree C. This is primarily due to the glass transition temperature of the hard segment ($T_gHS$) and the annealing endotherm that appears at 20–30 \textdegree C higher than the annealing temperature\cite{Saiani2007}, which was at 50 \textdegree C in this paper. This annealing endotherm is often related to the onset endothermic peak of short-ordered crystallites of the hard segment\cite{Kong2019,Puentes-Parodi2019,Verbelen2017}.
 Further heating resulted in a broad transition peak that comprises two or more hidden peaks, as shown in Fig~\ref{DSC_peaks}. These hidden peaks are attributed to the long-range hard segment with different morphologies. The final melting enthalpy peaks correspond to the long-range-ordered crystallites of the hard segment, which require more energy to break down and melt than previous peaks of melting enthalpy. The total melting enthalpy ($\Delta H_{T}$) is the sum of the areas under $\Delta H_{1}$,$\Delta H_{2}$, and $\Delta H_{3}$, which refer to melting enthalpy of micro-crystalline hard domains with highly ordered long-range hard segments, disordered long-range hard segments, and para-crystalline with short-range hard segments, respectively\cite{razeghi2018tpu}.

Upon comparing the $\Delta H_{T}$ values shown in Fig~\ref{enthalpy} for pristine samples (unaged) with different GMB volume fractions, it is observed that the TPU with 40\% GMB has the highest value, followed by TPU with 20\% GMB, and then Neat TPU. This can be attributed to the presence of GMBs in the powder mix can that potentially affect laser processing in selective laser sintering (SLS). The GMBs and TPU have different optical and thermal properties, which can result in non-homogeneous heat supplied by the laser to melt the TPU and GMB powder mixture. This contributes to more energy being supplied to the TPU within the GMB mixture, which increases the different morphologies of hard phases in the material. 


After 90 days of moisture exposure (Fig~\ref{enthalpy}), the total enthalpy increased drastically for neat TPU, followed by TPU with 20\% GMB, and only a marginal increase in TPU with 40\% GMB. One important reason for this is that the polar water molecules that diffused into the material attacked the original hydrogen bond (NH---C=O bond) with higher fracture energy\cite{Xu2021} to form weak hydrogen bonds with free carbonyl groups or amine groups. Upon desorbing the material at 50 °C for 24 hours, these hard segments dissociate, reorganize, and form different morphologies of crystallites. At 90 days, the water does not influence the polyester since hydrolysis-induced chain scission tends to occur much later.
At 160 days of moisture exposure, hydrolysis of the soft segments dominates. Chain scission of the long-chain polyol occurs at an exponential rate\cite{Bardin2020}, producing more short-chain diol. This diol then forms bonds with the urethane hard segments, spreading the hard segments homogeneously throughout the multiphase system. As a result, the total enthalpy decreases for all samples.

\subsection{Influence of moisture and GMB volume fraction on viscoelastic properties of syntactic foam}\label{viscoelastic_characterization}
Dynamic Mechanical Analysis (DMA) has been crucial in obtaining insightful data regarding the viscoelastic properties of materials. These properties, including storage modulus, loss modulus, and loss tangent, have been explored in the current research with regard to temperature, GMB volume fraction, and moisture exposure duration. Additionally, the identification of the glass transition temperature was achieved through the examination of the loss modulus peak.

The storage modulus (E') measures a material's ability to store elastic energy when subjected to deformation. In contrast, the loss modulus (E"), also known as viscous modulus, measures the energy dissipated in the form of heat when a material is subjected to deformation. The ratio of the loss modulus to the storage modulus is known as the loss tangent (tan$\delta$). This dimensionless parameter provides information about the damping characteristics of a material.

\begin{figure}[H]
 \centering
  \includegraphics[width=16cm]{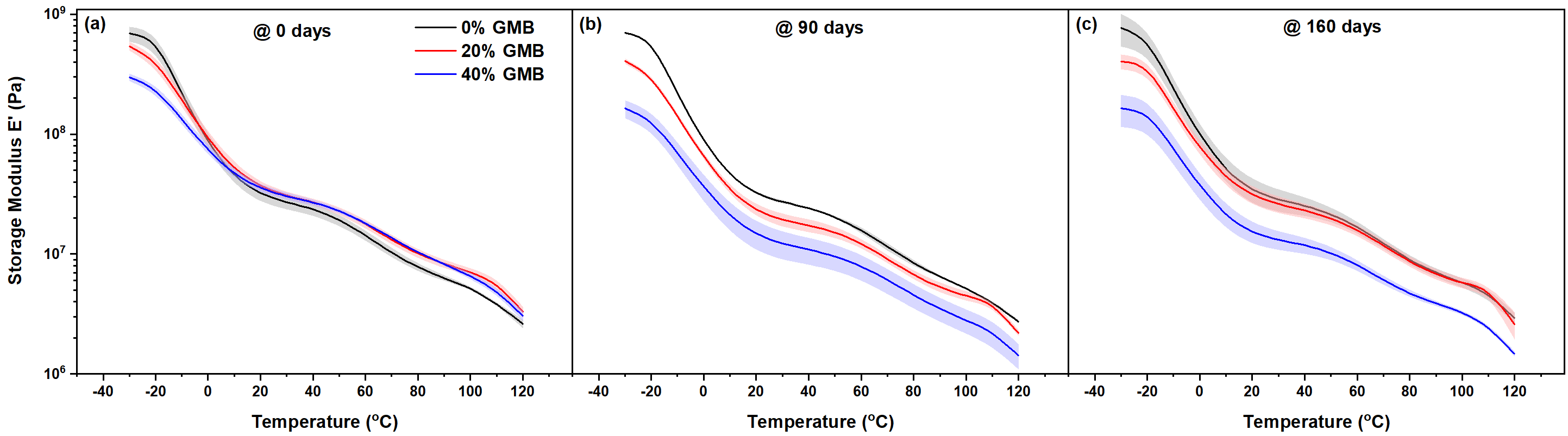} 
  \caption{Storage modulus vs. temperature plot for neat TPU and TPU with 20\% and 40\% GMB reinforcement at different exposure times of 0 days, 90 days, and 160 days. }
  \label{modulus-storage}
\end{figure}

\begin{figure}[H]
 \centering
  \includegraphics[width=16cm]{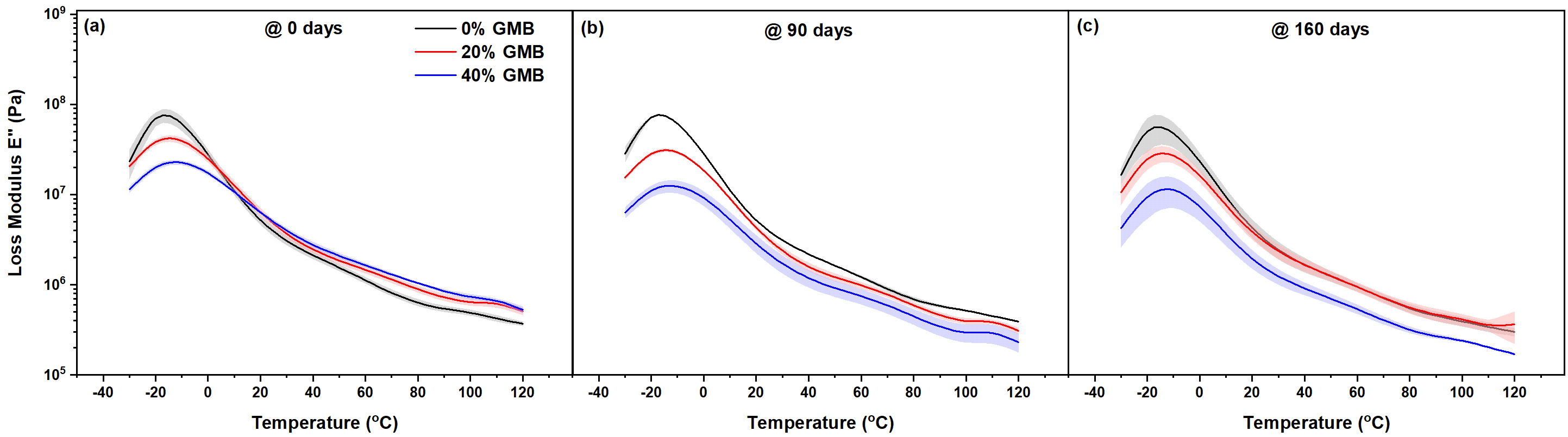} 
  \caption{Loss modulus vs. temperature plot for neat TPU and TPU with 20\% and 40\% GMB reinforcement at different exposure times of 0 days, 90 days, and 160 days. }
  \label{modulus-loss}
\end{figure}

\begin{figure}[H]
 \centering
  \includegraphics[width=16cm]{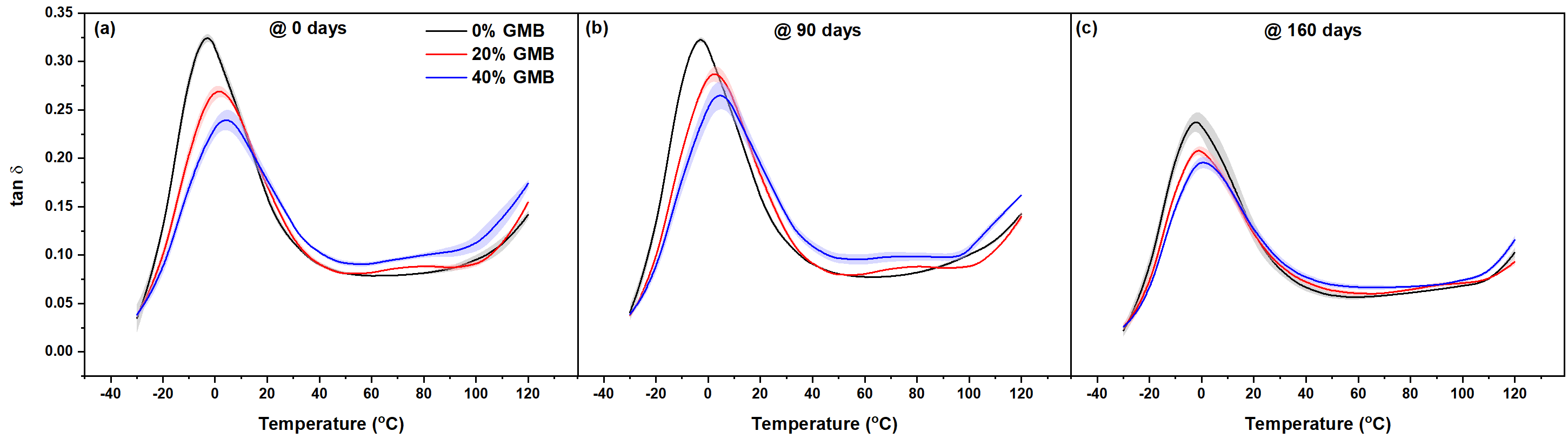}
  \caption{Loss tangent vs. temperature plot for neat TPU and TPU with 20\% and 40\% GMB reinforcement at different exposure times of 0 days, 90 days, and 160 days.}
  \label{tan}
\end{figure}

 Storage modulus, loss modulus, and tan$\delta$ are plotted against temperature at different moisture exposure times as shown in Fig~\ref{modulus-storage}, \ref{modulus-loss}, and \ref{tan}, respectively. 
 Temperature significantly influences the viscoelastic behavior of polymers and polymer composites, manifesting distinct regions such as the sub-Tg or glassy region, rubbery or leathery region, and terminal region. The glassy region is commonly observed at very low temperatures (below the glass transition temperature) or high frequencies, where the tightly packed polymer chains give rise to viscoelastic properties driven by local molecular motion. With increasing temperature, the rubbery or leathery region is entered, characterized by a considerable relaxation of the polymer chains via segmental motion and a peak in loss modulus and damping factor, commonly called the glass transition temperature or alpha peak. The onset of polymer melting denotes the terminal region, typically associated with high temperatures or low frequencies, where the polymer chains without any cross-links slide over each other and exhibit viscous behavior when it is thermally activated by high temperature\cite{lakes_2009}.
 As temperature increases, all samples' storage and loss modulus decreases. This phenomenon is often attributed to the relaxation of polymers, which is frequently associated with molecular rearrangements. In general, an increase in temperature leads to an expansion of the free volume of chain segments, thereby enhancing the mobility of the polymer chains and resulting in a greater degree of conformational freedom, which, in turn, leads to a lower modulus\cite{Menard2008}.

 From Fig~\ref{modulus-storage}(a) and \ref{modulus-loss}(a), during the analysis of unaged samples, it is noted that the storage modulus experiences a decrease as the GMB volume fraction increases within the sub-$T_g$ region. The increase in GMB content reduces the number of polymer chains that undergo configurational changes contributing to viscoelasticity. The trend is similarly observed for the loss modulus. Conversely, beyond the glassy state, the transition or leathery region gradient in both the storage and loss modulus decreases with an increase in the volume fraction of GMB. This is due to stiff inclusions, which disrupt polymer chains and hinder their molecular mobility as the temperature increases, leading to a broad transition. The storage modulus of TPU with 20\% and 40\% GMB is slightly higher than that of neat TPU up to 90 \textdegree C. Likewise, the loss modulus of a TPU with GMB decreases with an increase in GMB volume fraction at temperatures below the glass transition temperature ($T_g$). At this regime, the free volume of polymer reduces and, thereby, polymer chain mobility, eventually decreasing the friction between polymer chains. As the temperature increases above $T_g$, the friction between GMB and TPU becomes significant compared to polymer chain friction. This increases the energy dissipation, and so the loss modulus increases with an increase in GMB beyond $T_g$. The GMBs are not chemically bonded to the TPU, but the physical bonding due to processing is not degraded in the pristine specimens. The degree of interface degradation due to moisture will be explained next.

 As illustrated in Fig~\ref{modulus-storage}(b) and \ref{modulus-loss}(b), the storage and loss moduli exhibit a decreasing trend with an increase in the volume fraction of GMB for the entire experimental temperature range after 90 days of moisture exposure. This can be attributed to two main reasons. Firstly, the deterioration of interfacial adhesion between GMB and Thermoplastic Polyurethane (TPU) can occur. Secondly, the reduction in phase mixing in TPU with increasing GMB due to moisture leads to the formation of inhomogeneous phases with more short-range disordered hard segments and a different morphology of hard domains.

Based on the data presented in Fig~\ref{modulus-storage}(c), it can be observed that after 160 days of moisture exposure, the storage modulus of all samples decreases as the volume fraction increases before reaching room temperature. However, beyond room temperature, the difference in storage modulus between neat TPU and TPU with 20\% GMB gradually decreases as the temperature increases. This trend persists until it merges with the storage modulus of neat TPU at around 60 \textdegree C, remaining equivalent for the entire temperature range. Notably, the storage modulus of TPU with 40\% GMB remains lower than that of the other samples across the entire temperature range. This is primarily because the degree of phase mixing increases for 20\% and 40\% of GMB-reinforced samples after 160 days of moisture exposure. Prolonged exposure to moisture reduces soft segment chain length, giving rise to short and rigid polyol that bonds with the hard segment, thereby significantly increasing the storage modulus of 20\% GMB samples. However, the storage modulus for 40\% GMB samples does not improve despite increased phase mixing. This is because, for both 20\% and 40\% GMB samples, the GMBs are completely debonded, and the mechanical property contribution is primarily due to the polymer itself. In the case of 40\% GMB samples, the polymer contribution is far less than that of the other two types of samples. A similar trend is observed in loss modulus after 160 days of moisture exposure as shown in Fig~\ref{modulus-loss}(c).

Upon comparing the effect of moisture aging on each sample with its respective unaged sample, it was observed that the storage modulus of neat TPU remained relatively unchanged after 90 days of moisture exposure but exhibited a slight increase after 160 days. The loss modulus of neat TPU followed a similar trend for the first 90 days, up to 50 \textdegree C, but thereafter exhibited a slight increase, followed by a significant decrease after 160 days of moisture exposure. In contrast, the storage modulus of TPU with 20\% GMB decreased after 90 days of moisture exposure when compared with its unaged counterpart. However, a tremendous increase in storage modulus was observed after 160 days of exposure. Similarly, the loss modulus of TPU with 20\% GMB decreased significantly after 90 days of moisture exposure but exhibited a slight decrease after 160 days until reaching room temperature, after which it remained consistent with the 90-day aged sample. When comparing the storage modulus of TPU with 40\% GMB with its unaged samples, it was found to decrease drastically after 90 days of exposure but remained relatively unchanged after 160 days of moisture exposure. In contrast, the loss modulus of TPU with 40\% GMB exhibited a decrease with increasing moisture exposure time. An essential factor contributing to this is the higher quantity of reaction sites, including hard-hard segment, hard-soft segment, hard segment, and soft segment hydrolysis, for neat TPU. However, this amount decreases with an increase in the volume fraction of GMB, while the amount of water molecules (reactant) remains constant for all three distinct specimens. In comparison to neat TPU, the hydrolysis of the soft segment leads to exponential chain scission on both TPU with 20\% and 40\% GMB at a much faster pace.

 From Fig~\ref{tan}, the results show that adding more GMB to TPU decreases peak damping. Specifically, samples with 20\% and 40\% GMB volume fractions have peak damping decreases of 16.5\% and 26.17\%, respectively, compared to neat TPU. Furthermore, the study found that the peak damping of neat TPU remained relatively stable after 90 days of exposure but decreased by 26\% after 160 days, compared to unaged TPU. Samples with 20\% GMB increased by 6.37\% after 90 days of exposure but decreased by 22.96\% after 160 days compared to unexposed samples. In the case of TPU with 40\% GMB, peak damping increased by 10.82\% after 90 days of exposure but decreased by 17.43\% after 160 days when compared with unaged 40\% GMB samples.

\subsubsection*{Adhesion factor}\label{adhesion chapter}
The dynamic mechanical analysis provides insightful data to understand the degree of adhesion in heterogeneous mediums. The damping factor (tan $\delta$) determines the adhesion strength between GMB filler and TPU matrix using the adhesion factor proposed by Kubat et al.\cite{Kubat}. The tan$\delta_f$ is assumed to be 0 since the damping of rigid filler is extremely small.

\begin{equation}
    tan\: \delta_c = V_f \; tan\: \delta_f + V_i \; tan\: \delta_i + V_m \; tan\: \delta_m
\end{equation}
\begin{equation}
    \frac{tan\: \delta_c}{tan\: \delta_m} = (1-V_f)(1+A)
\end{equation}
where,

\begin{equation}
    A = \frac{1}{1-V_{f}}\frac{tan\: \delta_c}{tan\: \delta_m} - 1
\end{equation}

Let us consider the Adhesion factor as $A_0$ and $A_d$, when we utilize $tan \delta_m$ of neat TPU that is unaged and aged samples for specified, respectively. This is to differentiate accelerated adhesion that happens in TPU hard and soft segments for neat TPU and TPU with GMB incorporation. In the context of GMB filler and TPU matrix interface, a perfect bond results in diminished macromolecular mobility near the interface, reducing interface damping and adhesion factors.

\begin{figure}[h!]
\centering
\subfigure[]{
\includegraphics[width=7cm]{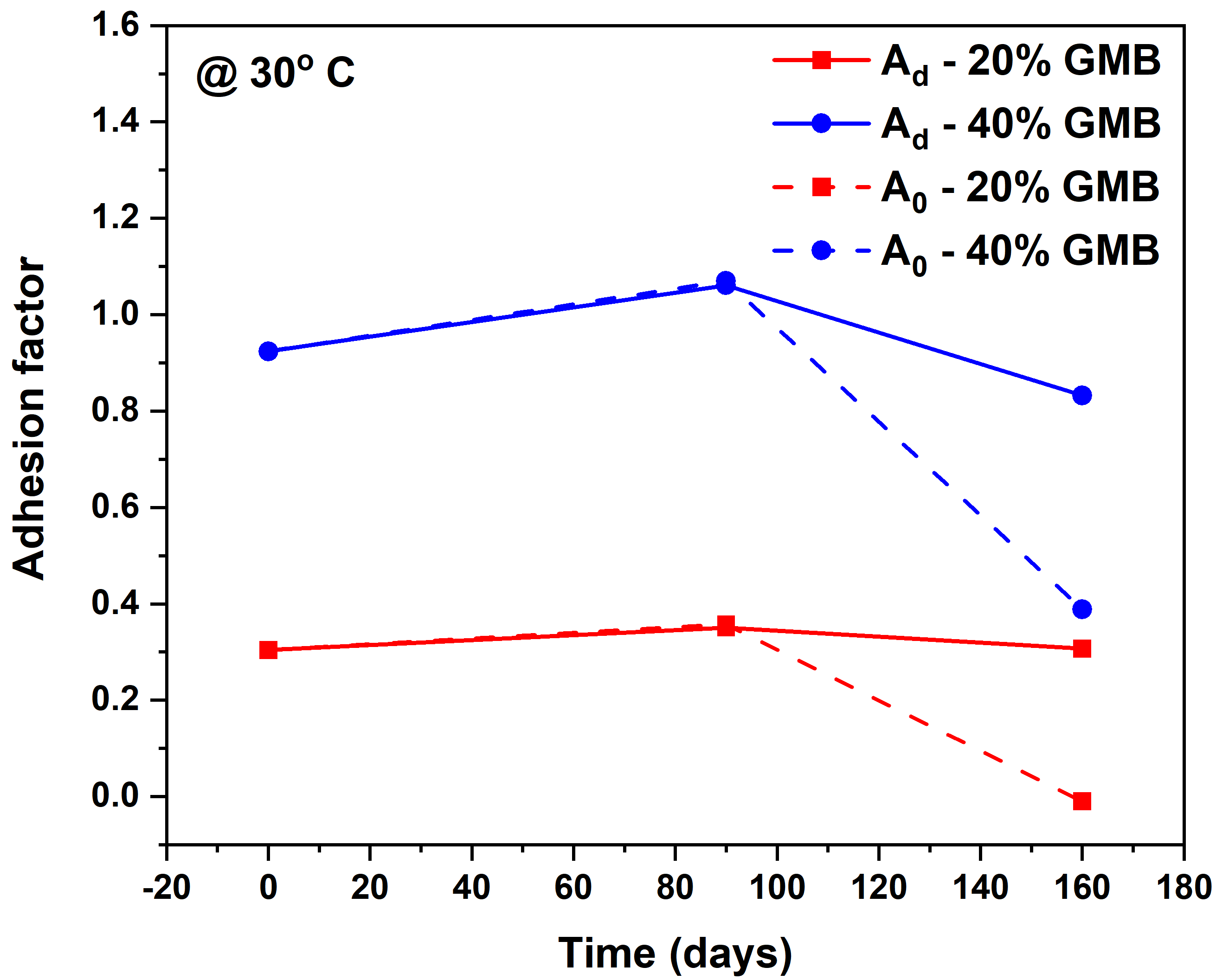}\label{adhesion}
}
\hspace{0.2in}
\centering
\subfigure[]{
\includegraphics[width=7cm]{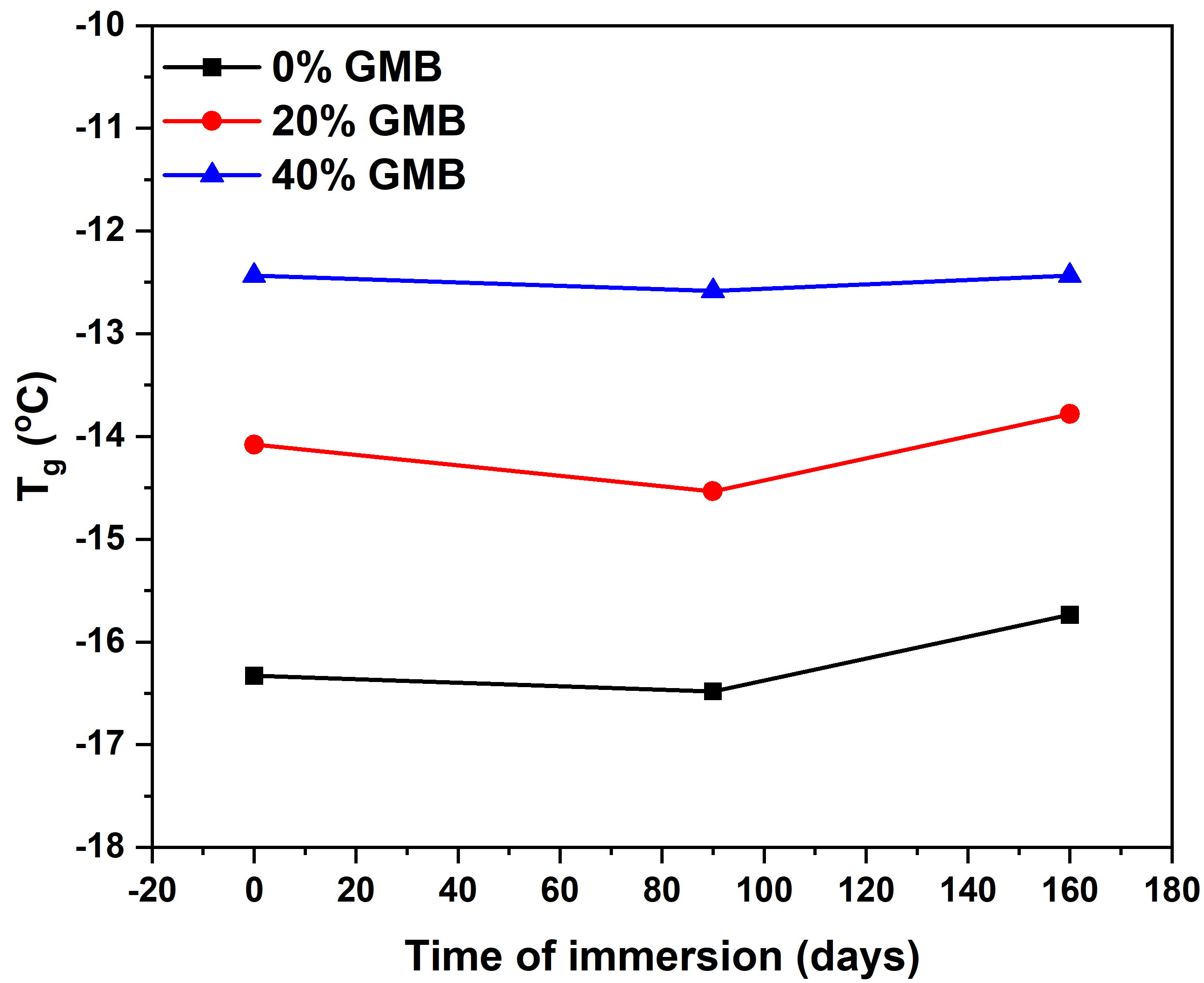}\label{Tg}
}
\caption{(a) Adhesion factor as a function of moisture exposure time and GMB volume fraction at 30\textdegree C and (b) Glass transition temperature from the peak of loss modulus as a function of GMB volume fraction and time of moisture immersion}
\end{figure}

When the adhesion factor approaches zero, it indicates perfect bonding. However, as the adhesion factor increases, it denotes a deterioration in the interface between two distinct media. Fig~\ref{adhesion} illustrates the adhesion factor of TPU with 20\% and 40\% GMB at 30 \textdegree C for varying moisture exposure times. Upon comparing the unaged samples, TPU with 20\% GMB has a lower adhesion factor than TPU with 40\% GMB. This denotes that an increase in the volume fraction of GMB tends to have a higher fraction of deteriorated interface. After 90 days of exposure, the interface between GMB and TPU further deteriorates due to dissimilar moisture swelling of GMB and TPU. On the other hand, after 160 days of moisture exposure, there is a slight decrease in the adhesion factor of TPU with 40\% GMB, while there is no change in the adhesion factor for TPU with 20\% GMB. This slight variation may be attributed to adhesion between the hard and soft segments or hard to hard-segments in TPU. To distinguish this, the adhesion factor that accounts for enhanced adhesion in TPU copolymer uses the damping factor of the respective aged neat TPU to calculate the adhesion factor. Thus, at 160 days of moisture exposure, the chain scission of the soft segment creates stiffer and shorter chains that bond with the hard segment to form increased adhesion.

The loss modulus peak in DMA is used here to quantify the glass transition temperature as transitions are closely tied to mechanical properties. Fig~\ref{Tg} shows the influence of moisture and GMB volume fraction on the glass transition temperature, which increases with an increase in the GMB volume fraction. This is because the mobility of the polymer chain that occurs in the transition is often hindered by the presence of GMB. The glass transition temperature of all samples decreases slightly after 90 days of moisture exposure and then increases after 160 days when compared to pristine samples. One of the main reasons for this behavior is that the hard segment present in the soft domain decreases due to the decreased degree of phase mixing after 90 days, whereas after 160 days of moisture exposure, the degree of phase mixing increases, as explained in the section~\ref{chemical_change}.


\section{Conclusion}\label{conc}
This paper aims to explore the long-term moisture-induced degradation in TPU and TPU-based syntactic foam, along with its associated chemical and microphase morphology changes. To this end, samples were immersed in water and tested at two different time points: 90 days and 160 days. Additionally, the study aims to shed light on the manifestation of these changes in viscoelastic properties of TPU and TPU-based syntactic foam. 

    The alteration of the total melting enthalpy of TPU and TPU-based syntactic foam after prolonged moisture exposure provided insights into various degradation mechanisms linked to the internal morphology of TPU and TPU-based syntactic foams. After 90 days of moisture exposure, the total melting enthalpy for all types of specimens increases significantly due to the cleavage of the original hydrogen bond, followed by the aggregation of polar hard segments to form different morphologies of hard domains. However, after 160 days of exposure, the long-chain polyol (soft segment) chain scission increases exponentially. The formation of new bonds between hard and soft segments creates a homogenous mix of the same, leading to a decrease in the total melting enthalpy. The study also finds similar behavior using FTIR, where the hard domain increases after 90 days of exposure while decreases (Degree of Phase Mixing increases) after 160 days of moisture exposure.

    The addition of GMB and moisture exposure alters the polymer's viscoelastic properties. At sub-Tg, an increase in GMB volume fraction leads to a decrease in storage modulus because the rise in GMB content reduces the number of polymer chains that undergo configurational changes contributing to viscoelasticity. However, after $T_g$, the storage modulus increases with an increase in GMB content because the free movement of polymer chains is often hindered by the presence of GMB. After 90 days of exposure, the TPU-GMB exhibits significant deterioration, accompanied by localized aggregation of hard segments that result in an inhomogeneous mixture. On the other hand, after 160 days of exposure, the storage modulus of TPU with 20\% GMB increases significantly, primarily due to the uniform distribution of soft and hard segments in TPU itself. This is because the GMB is debonded after 90 days of exposure, allowing the hard segments to be entrapped in soft domains, acting as physical crosslinks or rigid fillers. Additionally, the loss modulus decreases significantly due to reduced internal friction.
    
To conclude, this work not only aids in comprehending the underlying degradation mechanisms caused by moisture and temperature in TPU and TPU-based syntactic foam, but it also sheds light on the possibilities of tailoring TPU and TPU-based syntactic foam at different length scales to achieve desirable viscoelastic properties.
    


\section*{Author Contributions}

{\bf Sabarinathan P Subramaniyan:} Conceptualization, Methodology, Formal Analysis, Visualization, Investigation, Writing - Original Draft. {\bf Pavana Prabhakar:} Conceptualization, Methodology, Writing - Original Draft, Visualization, Verification, Supervision, Project Administration, Funding Acquisition. 


\section*{Funding}

The authors would like to acknowledge the support of the National Science Foundation (NSF) CAREER Award \# 2046476 through the Mechanics of Materials and Structures (MOMS) Program for conducting the research presented here.


\section*{Acknowledgment}
This research was partially supported by the University of Wisconsin - Madison College of Engineering Shared
Research Facilities and the NSF through the Materials Science Research and Engineering Center (DMR-1720415)
using instrumentation provided at the UW - Madison Materials Science Center. 
 


\bibliographystyle{unsrt}
\bibliography{references}

\begin{thebibliography}{10}

\bibitem{Aurilia2011}
M~Aurilia, F~Piscitelli, L~Sorrentino, M~Lavorgna, and S~Iannace.
\newblock {Detailed analysis of dynamic mechanical properties of TPU
  nanocomposite: The role of the interfaces}.
\newblock {\em European Polymer Journal}, 47(5):925--936, 2011.

\bibitem{Bruckmoser2014}
K~Bruckmoser and K~Resch.
\newblock {Investigation of ageing mechanisms in thermoplastic polyurethanes by
  means of IR and Raman spectroscopy}.
\newblock {\em Macromolecular Symposia}, 339(1):70--83, 2014.

\bibitem{Deng1994}
Y~W Deng, T~L Yu, and C~H Ho.
\newblock {Effect of aging under strain on the physical properties of
  polyester–urethane elastomer}.
\newblock {\em Polymer Journal}, 26(12):1368--1376, 1994.

\bibitem{Osswald2012}
T~A Osswald and G~Menges.
\newblock {\em {Material Science of Polymers for Engineers}}.
\newblock Carl Hanser Verlag GmbH \& Co. KG, third edition edition, 2012.

\bibitem{Xu2021}
D~Xu, F~Liu, G~Pan, Z~G Zhao, X~Yang, H~C Shi, and S~F Luan.
\newblock {Softening and hardening of thermal plastic polyurethane blends by
  water absorbed}.
\newblock {\em Polymer}, 218(February):123498, 2021.

\bibitem{Xiu1992}
Y~Xiu, Z~Zhang, D~Wang, S~Ying, and J~Li.
\newblock {Hydrogen bonding and crystallization behaviour of segmented
  polyurethaneurea: effects of hard segment concentration}.
\newblock {\em Polymer}, 33(6):1335--1338, 1992.

\bibitem{Xiang2017}
D~Xiang, L~Liu, and Y~Liang.
\newblock {Effect of hard segment content on structure, dielectric and
  mechanical properties of hydroxyl-terminated butadiene-acrylonitrile
  copolymer-based polyurethane elastomers}.
\newblock {\em Polymer}, 132:180--187, 2017.

\bibitem{Walo2014}
M~Walo, G~Przybytniak, K~{\L}yczko, and M~Piatek-Hnat.
\newblock {The effect of hard/soft segment composition on radiation stability
  of poly(ester-urethane)s}.
\newblock {\em Radiation Physics and Chemistry}, 94(1):18--21, 2014.

\bibitem{Petcharoen2013}
K~Petcharoen and A~Sirivat.
\newblock {Electrostrictive properties of thermoplastic polyurethane elastomer:
  Effects of urethane type and soft-hard segment composition}.
\newblock {\em Current Applied Physics}, 13(6):1119--1127, 2013.

\bibitem{Yang2006}
B~Yang, W~M Huang, C~Li, and L~Li.
\newblock {Effects of moisture on the thermomechanical properties of a
  polyurethane shape memory polymer}.
\newblock {\em Polymer}, 47(4):1348--1356, 2006.

\bibitem{Boubakri2009}
A~Boubakri, K~Elleuch, N~Guermazi, and H~F Ayedi.
\newblock {Investigations on hygrothermal aging of thermoplastic polyurethane
  material}.
\newblock {\em Materials and Design}, 30(10):3958--3965, 2009.

\bibitem{Boubakri2010}
A~Boubakri, N~Haddar, K~Elleuch, and Y~Bienvenu.
\newblock {Impact of aging conditions on mechanical properties of thermoplastic
  polyurethane}.
\newblock {\em Materials and Design}, 31(9):4194--4201, 2010.

\bibitem{Puentes-Parodi2019}
A~Puentes-Parodi, L~A Santoro, I~Ferreira, A~Leuteritz, and I~Kuehnert.
\newblock {Influence of annealing on the permeation properties of a
  thermoplastic elastomer}.
\newblock {\em Polymer Engineering and Science}, 59(9):1810--1817, 2019.

\bibitem{Bardin2020}
A~Bardin, P~{Le Gac}, S~C{\'{e}}rantola, G~Simon, H~Bindi, and B~Fayolle.
\newblock {Hydrolytic kinetic model predicting embrittlement in thermoplastic
  elastomers}.
\newblock {\em Polymer Degradation and Stability}, 171:1--11, 2020.

\bibitem{mishra2015long}
A~Mishra, K~Seethamraju, J~Delaney, P~Willoughby, and R~Faust.
\newblock Long-term in vitro hydrolytic stability of thermoplastic
  polyurethanes.
\newblock {\em Journal of Biomedical Materials Research Part A},
  103(12):3798--3806, 2015.

\bibitem{choi2023degradation}
E~Y Choi and C~K Kim.
\newblock Degradation and lifetime prediction of thermoplastic polyurethane
  encapsulants in seawater for underwater acoustic sensor applications.
\newblock {\em Polymer Degradation and Stability}, 209:110281, 2023.

\bibitem{Gupta2014}
N~Gupta, S~E Zeltmann, V~C Shunmugasamy, and D~Pinisetty.
\newblock {Applications of polymer matrix syntactic foams}.
\newblock {\em Jom}, 66(2):245--254, 2014.

\bibitem{Afolabi2020}
L~O Afolabi, Z~M Ariff, S~Hashim, T~Alomayri, S~Mahzan, K~Kamarudin, and
  I~Muhammad.
\newblock {Syntactic foams formulations, production techniques, and industry
  applications: A review}.
\newblock {\em Journal of Materials Research and Technology},
  9(5):10698--10718, 2020.

\bibitem{Shahapurkar2018}
K~Shahapurkar, C~D Garcia, M~Doddamani, G~C {Mohan Kumar}, and P~Prabhakar.
\newblock {Compressive behavior of cenosphere/epoxy syntactic foams in arctic
  conditions}.
\newblock {\em Composites Part B: Engineering}, 135(June 2017):253--262, 2018.

\bibitem{amos2015hollow}
S~E Amos and B~Yalcin.
\newblock {\em Hollow glass microspheres for plastics, elastomers, and
  adhesives compounds}.
\newblock Elsevier, 2015.

\bibitem{Barber1977}
E~Barber, J~Nelson, and W~Beck.
\newblock {Improving Properties in Rigid Urethane Foams Using Glass Bubbles}.
\newblock {\em Journal of Cellular Plastics}, 13(6):383--387, 1977.

\bibitem{Hagarman1985}
J~A Hagarman, J~P Cunnion, and B~W Sands.
\newblock {Formulation and Physical Properties of Polyurethane Foam
  Incorporating Hollow Microspheres.}
\newblock {\em Proceedings of the SPI Annual Technical/Marketing Conference},
  pages 161--163, 1985.

\bibitem{Im2011}
H~Im, C~K Kim, and O~C Kwon.
\newblock {Fabrication of novel polyurethane elastomer composites containing
  hollow glass microspheres and their underwater applications}.
\newblock {\em ICCM International Conferences on Composite Materials}, pages
  7305--7312, 2011.

\bibitem{Tewani2022}
H~Tewani, M~Hinaus, M~Talukdar, H~Sone, and P~Prabhakar.
\newblock {Architected syntactic foams: a tale of additive manufacturing and
  reinforcement parameters}.
\newblock 2022.

\bibitem{yilgor2015critical}
I~Yilg{\"o}r, E~Yilg{\"o}r, and G~L Wilkes.
\newblock Critical parameters in designing segmented polyurethanes and their
  effect on morphology and properties: A comprehensive review.
\newblock {\em Polymer}, 58:A1--A36, 2015.

\bibitem{cheng2022review}
B~Cheng, W~Gao, X~Ren, X~Ouyang, Y~Zhao, H~Zhao, W~Wu, C~Huang, Y~Liu, X~Liu,
  H~Li, and R~Li.
\newblock A review of microphase separation of polyurethane: Characterization
  and applications.
\newblock {\em Polymer Testing}, 107:107489, 2022.

\bibitem{sahebi2019novel}
I~Sahebi~Jouibari, V~Haddadi-Asl, and M~Mirhosseini.
\newblock A novel investigation on micro-phase separation of thermoplastic
  polyurethanes: simulation, theoretical, and experimental approaches.
\newblock {\em Iranian Polymer Journal}, 28:237--250, 2019.

\bibitem{valim2022unraveling}
F~C~F Valim, G~P Oliveira, Gibran da~Cunha~V, L~B de~Paiva, C~Santillo,
  M~Lavorgna, and R~J~E Andrade.
\newblock Unraveling the impact of phase separation induced by thermal
  annealing on shape memory effect of polyester-based thermoplastic
  polyurethane.
\newblock {\em Journal of Applied Polymer Science}, 139(9):51723, 2022.

\bibitem{Saiani2007}
A~Saiani, A~Novak, L~Rodier, G~Eeckhaut, J~W Leenslag, and J~S Higgins.
\newblock {Origin of multiple melting endotherms in a high hard block content
  polyurethane: Effect of annealing temperature}.
\newblock {\em Macromolecules}, 40(20):7252--7262, 2007.

\bibitem{Kong2019}
Z~Kong, Q~Tian, J~Zhang, Rand~Yin, L~Shi, W~Ying, H~Hu, C~Yao, K~Wang, and
  J~Zhu.
\newblock {Reexamination of the microphase separation in MDI and PTMG based
  polyurethane: Fast and continuous association/dissociation processes of
  hydrogen bonding}.
\newblock {\em Polymer}, 185(October):121943, 2019.

\bibitem{Verbelen2017}
L~Verbelen, S~Dadbakhsh, M~{Van den Eynde}, D~Strobbe, J~P Kruth, B~Goderis,
  and P~{Van Puyvelde}.
\newblock {Analysis of the material properties involved in laser sintering of
  thermoplastic polyurethane}.
\newblock {\em Additive Manufacturing}, 15:12--19, 2017.

\bibitem{razeghi2018tpu}
M~Razeghi and G~Pircheraghi.
\newblock Tpu/graphene nanocomposites: Effect of graphene functionality on the
  morphology of separated hard domains in thermoplastic polyurethane.
\newblock {\em Polymer}, 148:169--180, 2018.

\bibitem{lakes_2009}
R~Lakes.
\newblock {\em Viscoelastic Materials}.
\newblock Cambridge University Press, 2009.

\bibitem{Menard2008}
K~P Menard.
\newblock {Dynamic Mechanical}.
\newblock {\em CRC Press, Taylor \& Francis Group}, Second Edition:1--35, 2008.

\bibitem{Kubat}
J~Kubát, M~Rigdahl, and M~Welander.
\newblock Characterization of interfacial interactions in high density
  polyethylene filled with glass spheres using dynamic‐mechanical analysis.
\newblock {\em Journal of Applied Polymer Science}, 39:1527--1539, 1990.

\end{thebibliography}

\end{document}